\documentclass[preprint,12pt]{elsarticle}

\usepackage{bm}
\usepackage{amsmath,amssymb,amsfonts,times}
\usepackage{url}

\usepackage[colorlinks=true, pdfstartview=FitV, linkcolor=red, citecolor=blue, urlcolor=blue]{hyperref}

\usepackage{graphicx}
\usepackage{epsfig}
\usepackage{slashed}
\usepackage{a4wide}
\usepackage{fancyhdr}
\usepackage{subfigure}

\usepackage{rotating}

\usepackage{pdflscape}

\allowdisplaybreaks[1]
\newcommand{\UNIT}[1]{\mbox{$\,{\rm #1}$}}
\newcommand{\MeV}{\UNIT{MeV}}

\newcommand{\fm}{\UNIT{fm}}

\newcommand{\be}{\begin{equation}}
\newcommand{\ee}{\end{equation}}
\newcommand{\ba}{\begin{eqnarray}}
\newcommand{\ea}{\end{eqnarray}}

\newcommand{\nld}{{\cal D}}

\newcommand{\nldl}{\overleftarrow{{\cal D}}}
\newcommand{\nldr}{\overrightarrow{{\cal D}}}

\newcommand{\Bla}{\Big<}
\newcommand{\Bra}{\Big>}

\newcommand{\panda}{$\overline{\mbox P}$ANDA~}
\newcommand{\Lagr}{\mathcal{L}}

\begin{document}

\begin{frontmatter}

\title{Toward relativistic mean-field description of $\overline{\text{N}}$-nucleus reactions}

\author{T.~Gaitanos $^{1,2}$, M. Kaskulov $^{2}$}
\address{$^{1}$ Department of Theoretical Physics, 
Aristotle University of Thessaloniki, GR-54124 Thessaloniki, Greece}
\address{$^{2}$ Institut f\"ur Theoretische Physik, Universit\"at Giessen,
             D-35392 Giessen, Germany}
\address{email: tgaitano@auth.gr}

\begin{abstract}
In this work we study the 
antinucleon-nucleus optical potential in the framework of the 
non-linear derivative (NLD) model with momentum dependent mean-fields. 
We apply the NLD model to interaction of 
antinucleons ($\overline{\text{N}}$) in nuclear matter and, in particular, 
to antiproton scattering on nuclei. In nuclear matter 
a strong suppression of the $\overline{\text{N}}$-optical potential at rest 
and at high kinetic energies is found and 
caused by the momentum dependence of relativistic mean-fields. 
The NLD results are consistent with known empirical $\overline{\text{N}}$-nucleus 
observations and agree well with antiproton-nucleus scattering data. 
This makes the NLD approach compatible with both, 
nucleon and antinucleon Dirac phenomenologies. 
Furthermore, in nuclear matter an effective mass splitting between nucleons and 
antinucleons is predicted.
\end{abstract}


\end{frontmatter}

\date{\today}

\section{\label{sec1}Introduction}

An interpretation of recent astrophysical 
observations on compact neutron stars~\cite{NS1,NS2} has driven 
the nuclear physics and astrophysics communities 
to detailed investigations 
of the nuclear equation of state (EoS) under conditions far beyond 
the ordinary matter~\cite{Lattimer14}. 
Theoretical and experimental studies on heavy-ion 
collisions~\cite{EoS,Fuchs,Aichelin,LeFevre,buss,gaithic,dani} 
over the last few decades concluded a softening of the EoS at high densities, 
which was compatible with phenomenological~\cite{pheno}, as well as with 
microscopic models~\cite{micro}. However, the recent 
observations of pulsars with masses of 
1.97 $\pm$ 0.04 $M_\odot$~\cite{NS1} and 2.01 $\pm$ 0.04 $M_\odot$~\cite{NS2} 
gave some controversial insights on the high-density EoS. These observations 
provide the upper limit for the neutron star mass by excluding a soft 
EoS at high baryon densities. 

On the other hand, the in-medium antinucleon-nucleon interactions are closely 
related to the high baryon density domain of the EoS. In highly compressed and cold 
matter the baryons can reach very high Fermi momenta. Such 
kinematical conditions can be also achieved 
in an antiproton-nucleus reaction for a short time before 
annihilation. In fact, in antiproton-nucleus collisions, where in rare events 
antiprotons penetrate deep inside the nucleus, significant 
compressional effects are expected~\cite{Larionov1}. Therefore, the 
study of antinucleon-interactions in nuclear medium can provide 
useful information and additional constraints on the nuclear EoS.

The traditional approach for the description of nuclear matter 
is a well established relativistic mean-field (RMF) 
theory~\cite{Duerr:1956zz,boguta,Serot:1984ey,Walecka:1974qa,Serot:1997xg}. 
However, as shown in Refs.~\cite{cass2,larionov,mishustin}, 
RMF cannot properly describe in-medium antinucleon interactions using the G-parity 
arguments only, which is supposed to be a symmetry of RMF. On the contrary, 
the microscopic models for the description of in-medium antinucleon potentials 
strongly rely on G-parity in construction of the bare antinucleon-nucleon 
interaction starting from the nucleon-nucleon potential~\cite{heidenbauer,abhf}. 
This is considered as a problem in RMF approach, therefore, different solutions 
have been proposed. For instance, in Ref.~\cite{larionov} a violation of G-parity 
in nuclear matter has been considered. In Ref.~\cite{nlde} another method was 
invented to resolve this issue by using energy dependent interactions. 
However, this approach has limitations in density and lacks the thermodynamic 
consistency.

In this work we address the issue why the conventional RMF models are not 
consistent with antiproton-nucleus Dirac phenomenology.  
Our studies are based on the non-linear derivative (NLD) model 
to RMF as formulated in Ref.~\cite{Gaitanos:2012hg}. The NLD model 
describes simultaneously the density dependence of the nuclear EoS, 
the momentum dependence of the proton-nucleus Schr\"{o}dinger-equivalent 
optical potential and 
the recent observations of neutron star masses. In this work we extend our 
studies from Ref.~\cite{Gaitanos:2012hg} and 
show that in a G-parity conserving NLD framework the in-medium proton 
and antiproton optical potentials can be consistently reproduced in agreement 
with empirical data. We further compare the NLD calculations with scattering data. 
It is demonstrated, in particular, that the NLD 
model can describe the antiproton-nucleus reaction cross section data fairly well. 
Furthermore, we discuss other effects related to the behavior of antinucleons 
in nuclear matter and predict a considerable in-medium splitting between 
nucleon and antinucleon effective masses. The relevance 
of our results for upcoming experiments at FAIR is discussed.

\section{\label{sec2}The NLD Model}

At first, we briefly discuss the NLD model and for additional details we refer 
to Refs.~\cite{Gaitanos:2012hg,Gaitanos:2009nt,Gaitanos:2011yb}. 
The NLD approach is based on a field theoretical formalism 
of relativistic hadrodynamics (RHD). The NLD  Lagrangian
\begin{align}
\Lagr = \Lagr_{Dirac} + \Lagr_{mes} + \Lagr_{int}
\end{align}
contains the free 
Lagrangians for the Dirac field $\Psi$ and for the $\sigma$, $\omega$ and 
$\rho$ meson fields. The essential difference to the conventional RHD 
shows up in the NLD interaction Lagrangian~\cite{Gaitanos:2012hg}
\begin{align}
{\cal L}_{int}=\sum_{m=\sigma,\omega,\rho}{\cal L}_{int}^{m}
\end{align}
with
\begin{equation}
{\cal L}_{int}^{m} = \frac{g_{m}}{2}
	\left[
	\overline{\Psi}
	\, \nldl \Gamma_{m}
	\Psi\varphi_{m}
	+\varphi_{m}\overline{\Psi}
	\, \Gamma_{m}\nldr
	\Psi
	\right]
\label{NLD}
\end{equation}
where 
$\varphi_{m}$ stands for the scalar-isoscalar $\sigma$, vector-isoscalar 
$~\omega^{\mu}$ and vector-isovector $\vec{\rho\,}^{\mu}$ 
meson fields with $m=\{\sigma,~\omega,~\rho\}$, with obvious 
notations for couplings $g_{m}$ and vertices 
$\Gamma_{m}=\{ \UNIT 1,~\gamma^{\mu},~\vec{\tau}\gamma^{\mu} \}$ 
($\vec{\tau}$ are the isospin matrices). 

In Eq.~(\ref{NLD}) $\nldl,~\nldr$ are non-linear derivative 
operators, see Ref.~\cite{Gaitanos:2012hg}. They generate 
(infinite) series of higher-order derivative terms in the generalized 
Euler-Lagrange equations of motion. 
This series of terms lead to a momentum (energy) cut-off, which regulates 
the high momentum behavior of relativistic mean-fields in the interaction 
Lagrangian. The important feature 
of NLD model is that, independent of the functional form of the non-linear 
operators, all the higher-order derivative interactions can be 
resummed exactly in the RMF approximation.

The in-medium RMF interactions are obtained from the 
equations of motion for all degrees of freedom. For the generalized functional 
in Eq.~(\ref{NLD}) the field-theoretical formalism has been developed 
in~\cite{Gaitanos:2012hg} 
and is applied in this work. In the RMF approximation 
the Dirac equation for nucleons ($N_{i}$) forming the nuclear matter reads 
\begin{eqnarray}
\left[
	\gamma_{\mu}(i\partial^{\mu}-\Sigma^{\mu}_{vi}) - 
	(m-\Sigma_{s})
\right]\Psi_{N_{i}} & = & 0
\,. \label{Dirac_p}
\end{eqnarray}
For antinucleons ($\overline{N}_{i}$) interacting with nuclear matter 
the Dirac equation is derived from the requirement of G-parity invariance
\begin{eqnarray}
\left[
	\gamma_{\mu}(i\partial^{\mu}-\overline{\Sigma}^{\mu}_{vi}) - 
	(m-\Sigma_{s})
\right]\Psi_{\overline{N}_{i}} & = & 0
\label{Dirac_pbar}
\,.
\end{eqnarray}
The nucleon and antinucleon vector self-energies are given by
\begin{align}
\Sigma^{\mu}_{vi} & =  g_{\omega}\omega^{\mu}{\cal D}
+g_{\rho} \tau_{i}  \rho^{\mu}{\cal D} ~~~~,~
\label{Sigmav_nm}
\\
\overline{\Sigma}^{\mu}_{vi} & =  -g_{\omega}\omega^{\mu}{\cal D}
+g_{\rho} \tau_{i}  \rho^{\mu}{\cal D} 
\label{Sigmas_nm}
\end{align}
where now $\tau_{i}=+1$ for protons ($i=p$) and $\tau_{i}=-1$ 
for neutrons ($i=n$). Note the opposite signs in the isoscalar-vector 
interactions in Eqs.~(\ref{Sigmav_nm}) and~(\ref{Sigmas_nm}) 
between nucleons and antinucleons 
and that both, the vector and scalar components of the self-energy, 
depends explicitly on the particle momentum $\vec{p}$ through the NLD regulator 
function $\nld=\nld(\vec{p\,})$ in momentum representation. 
The scalar self-energy in Eqs.~(\ref{Dirac_p}) 
and~(\ref{Dirac_pbar}) reads as 
\begin{align}
\Sigma_{s}  =  g_{\sigma}\sigma\nld
\,.
\end{align}
Though the scalar self-energies between nucleons and 
antinucleons are formally the same, the kinetic energy dependence will generate 
a difference between them. Therefore, we distinguish them by different symbols 
in the following.

The single particle energies $E$ for nucleons and antinucleons 
are obtained from the corresponding in-medium 
mass-shell conditions 
\begin{equation}
E_{N_{i}} = \sqrt{p^{2}+m^{*2}}+\Sigma_{vi}^{0}
\,,
\label{mass-shel-N}
\end{equation}
\begin{equation}
E_{\overline{N}_{i}} = \sqrt{p^{2}+\overline{m}^{*2}}+\overline{\Sigma}_{vi}^{0}
\;, \label{mass-shel-Nbar}
\end{equation}
with the in-medium (effective or Dirac) masses of nucleons and antinucleons, 
given by $m^{*}=m-\Sigma_{s}$ and 
$\overline{m}^{*}=m-\overline{\Sigma}_{s}$, respectively. 
They depend explicitly on particle momentum. 

In nuclear matter the NLD equations of motion for the meson-fields are reduced 
to standard RMF algebraic equations 
\begin{align}
m_{\sigma}^{2}\sigma + \frac{\partial U(\sigma)}{\partial\sigma} = & g_{\sigma}
\sum_{i=p,n}\,\Bla \overline{\Psi}_{i}{\cal D}\Psi_{i}\Bra
~,\\
m_{\omega}^{2}\omega = & 
g_{\omega}
\sum_{i=p,n}\,\Bla \overline{\Psi}_{i} \gamma^{0}{\cal D}\Psi_{i}\Bra
~,\\
m_{\rho}^{2}\rho = & 
g_{\rho}
\sum_{i=p,n}\,\tau_{i}\Bla \overline{\Psi}_{i} \gamma^{0} {\cal D}\Psi_{i}\Bra
\label{mesonsNM}
\end{align}
where the various quantities and Lorentz-densities are given 
in Ref.~\cite{Gaitanos:2012hg}. The densities on the r.h.s in above equations depend 
explicitly on the particle momentum. The 
meson-field equations show a similar form as in standard RMF, however, the 
essential difference between the NLD approach and conventional RMF appears 
in the various Lorentz-sources of the meson fields, which get regulated at 
high momenta. 

The way one regulates the high-momentum behavior of fields is 
phenomenological. In Ref.~\cite{Gaitanos:2012hg} a simple monopole form 
\begin{equation}
\nld = \frac{\Lambda^2}{\Lambda^2 + \vec{p}^{\,2}}
\,,
\label{cutoff}
\end{equation}
has been used where $\Lambda$ is a cut-off parameter. It has been found 
that Eq.~(\ref{cutoff}) was able to accommodate the saturation properties 
of nuclear matter and, at the 
same time, the energy dependence of the Schr\"{o}dinger-equivalent optical 
potential. Note, that in the limit $\Lambda\to\infty$ one has 
$\nld \to 1$, such that the original RHD Lagrangian is recovered. 
Meanwhile in Refs.~\cite{yanjun1,typel} the NLD model has been 
tested with different choices of the cut-off function. 

\begin{figure}[t]
\begin{center}
\unitlength1cm
\begin{picture}(12.,10.0)
\put(-0.5,0.0){
\makebox{\includegraphics[clip=true,width=0.8\columnwidth,angle=0.]
		 {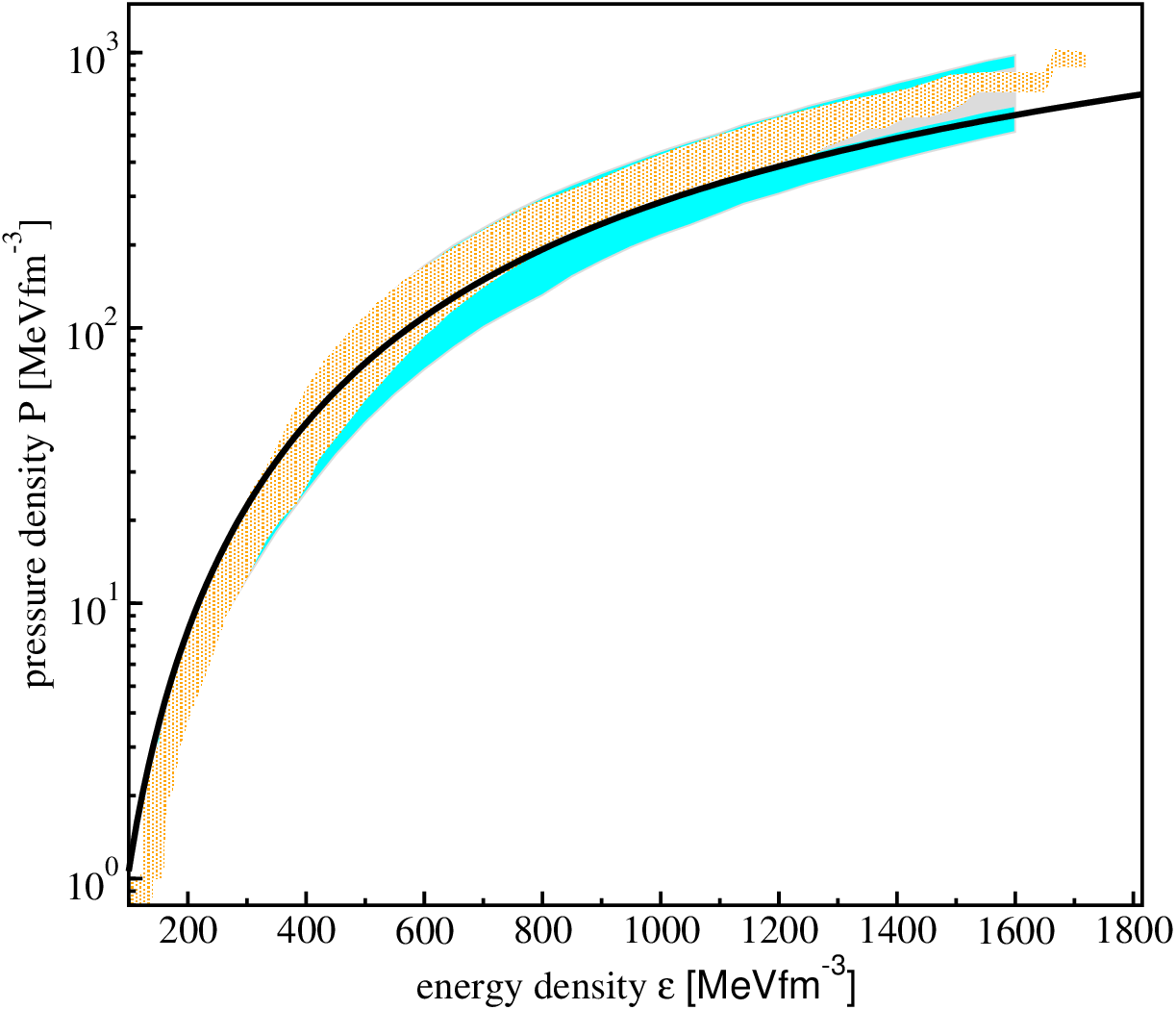}
		 }
}
\put(4.75,1.5){
\makebox{\includegraphics[clip=true,width=0.42\columnwidth,angle=0.]
		 {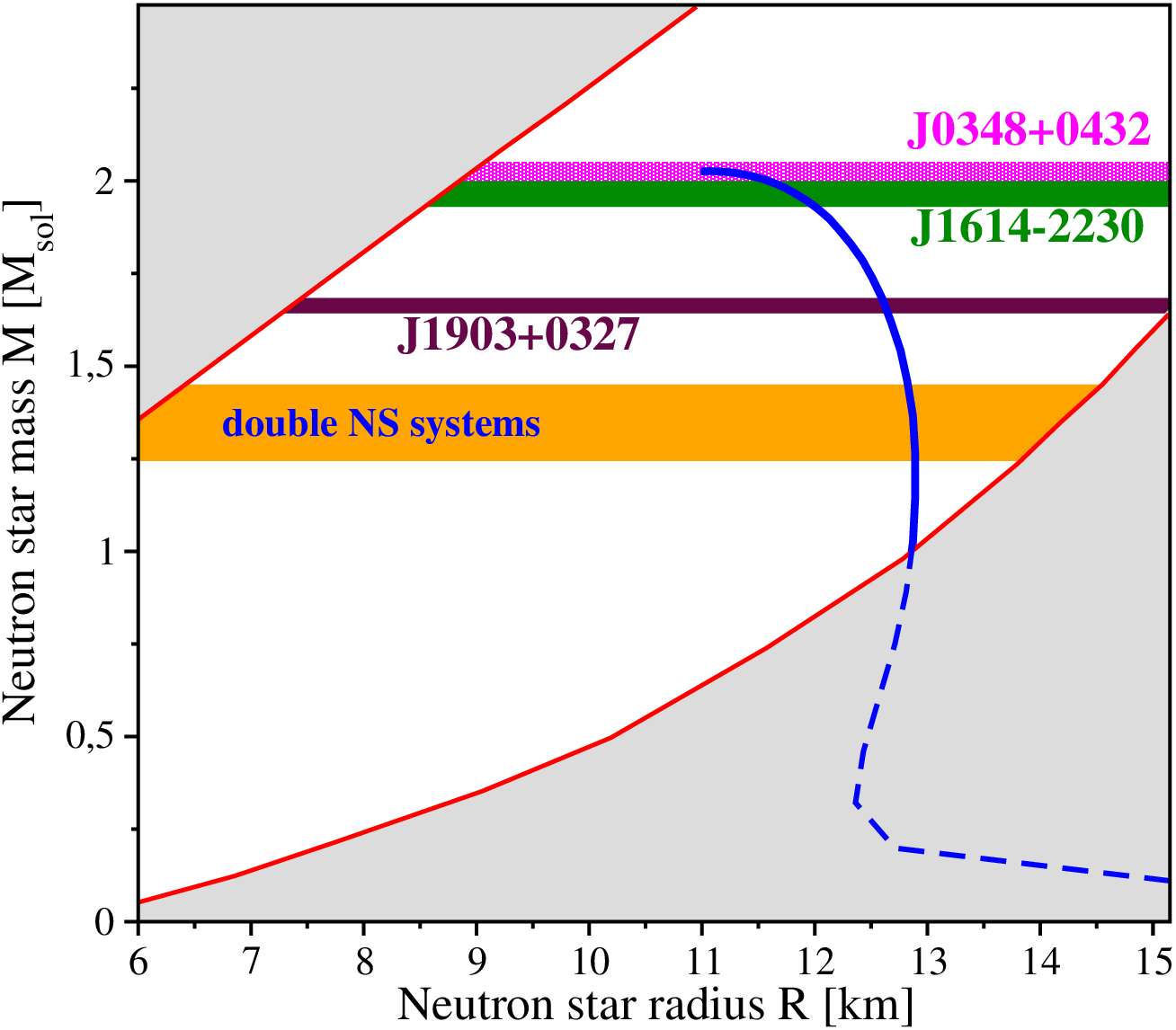}
		 }
}
\end{picture}
\caption{(Main panel) Pressure P as function of the energy density 
$\varepsilon$ in the NLD model. The shaded areas are the results of empirical 
studies~\cite{Lattimer14,Steiner10}. 
(Inserted panel) Mass-radius relation for neutron stars as 
calculated in NLD model, together 
with results from astrophysical observations and constraints:  
the four horizontal shaded bands refer to measurements 
from double neutron star (NS) systems~\cite{Lattimer:2004pg,Lattimer:2006xb}, 
from the pulsars PSR J1903+0327~\cite{Freire:2010tf}, 
PSR J1614-2230~\cite{NS1} and PSR J0348+0432~\cite{NS2}. The other shaded areas 
bordered by thick curves indicate parameter space excluded by general 
relativity, causality (shaded area on the top-left) and rotational 
constraints (shaded area on the bottom-right)~\cite{Lattimer:2004pg,Lattimer:2006xb}. 
}
\label{Fig1}
\end{center}
\end{figure}
The NLD approach is thermodynamically consistent~\cite{Gaitanos:2012hg}. 
The parameters of the model (coupling constants and cut-off) were adjusted to 
the empirical 
properties of nuclear matter and to the momentum dependence of the in-medium 
potential at saturation. As discussed in detail in~\cite{Gaitanos:2012hg}, 
the NLD approach can be reliably 
extrapolated to very high densities. As an example, in Fig.~\ref{Fig1} we 
demonstrate our new calculations of the equation of state (EoS) 
(pressure versus energy density, 
main panel) for nuclear matter in $\beta$-equilibrium. The NLD calculation 
(solid curve) reproduces the 
empirical analyses from astrophysical studies (shaded areas) fairly 
well~\cite{Lattimer14,Steiner10}. Of particular interest 
are the constraints from neutron star measurements. This is shown in the inner 
part of Fig.~\ref{Fig1} in terms of the mass-radius relation of neutron stars, 
where the NLD result is compared with various astrophysical observations 
(double NS systems and J1903+0327), and in particular, with the recent measurements 
on binary pulsars~\cite{NS1,NS2}. 

The explicit momentum-dependence of the NLD mean-fields 
is essential 
for a simultaneous description of low and high density observations involving, 
in particular, the empirical results from Dirac-phenomenology on the 
Schr\"{o}dinger-equivalent optical potential. The latter is important for 
the extension of our studies to in-medium antinucleon interactions.

\section{\label{sec5}Results and discussion}

The parameters of the NLD model used here have already been fixed in 
Ref.~\cite{Gaitanos:2012hg} 
at the bulk properties of nuclear matter and in-medium proton optical potential. 
Therefore, we do not readjust any parameters in this work when extending the NLD 
model to the description of antinucleons in nuclear matter. We just use the 
G-parity arguments for the construction of the in-medium antinucleon interaction.  

\begin{figure}[t]
\begin{center}
\unitlength1cm
\begin{picture}(12.,8.0)
\put(-1.25,0.0){
\makebox{\includegraphics[clip=true,width=0.9\columnwidth,angle=0.]
		 {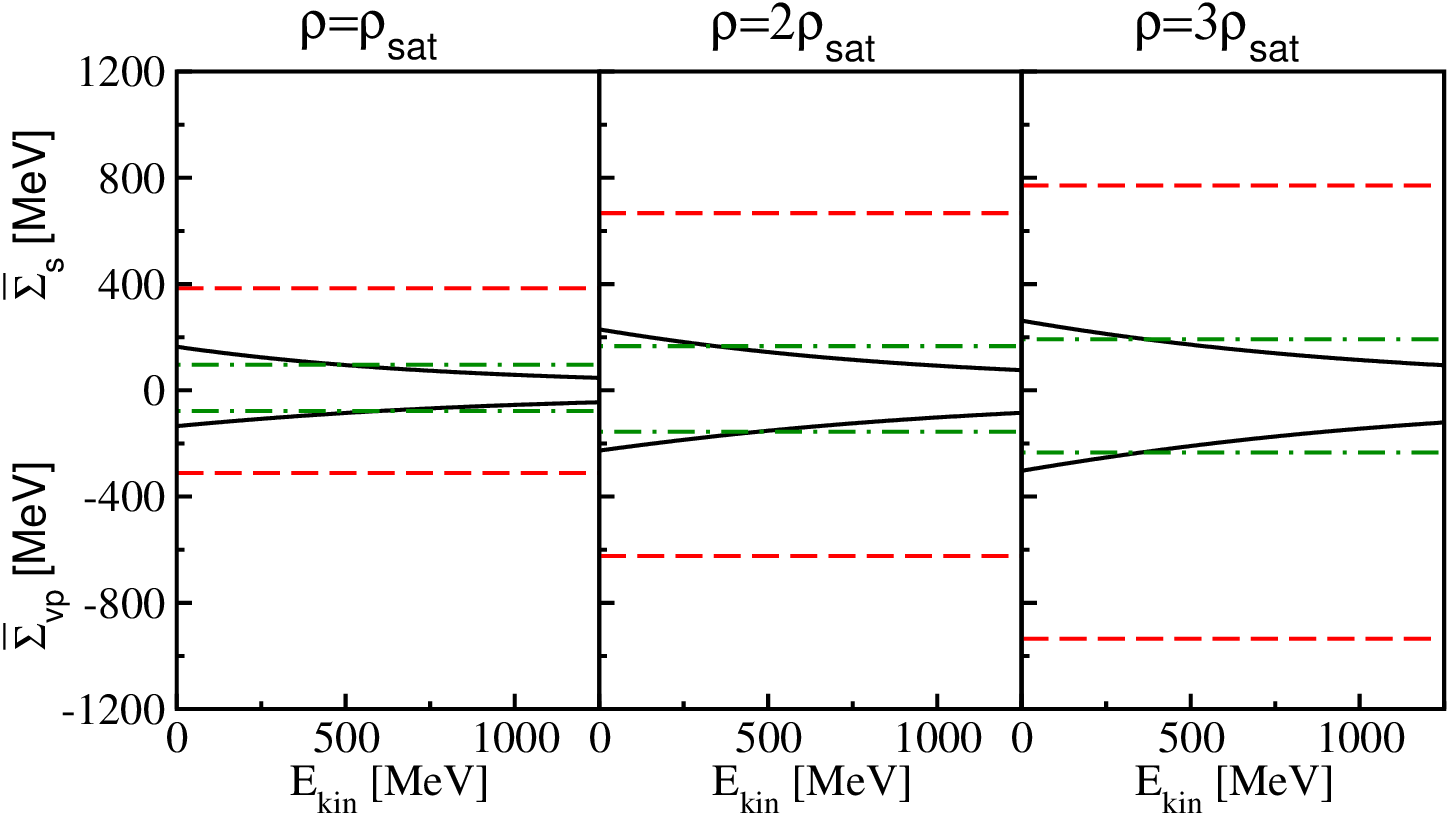}
		 }
}
\end{picture}
\caption{Kinetic energy dependence of the scalar and vector Lorentz-components 
of the antiproton self-energy in nuclear matter at densities of 
$\rho=\rho_{\rm sat}$ (left), $\rho=2\rho_{\rm sat}$ (middle) and
$\rho=3\rho_{\rm sat}$ (right)
using the NL3 model (dashed lines), the NL3 model with rescaled 
couplings with the factor $0.25$~\cite{larionov} (dash-dotted) 
and the NLD approach (solid lines). 
}
\label{Fig2}
\end{center}
\end{figure}
We start the discussion with the in-medium antiproton scalar and vector 
self-energies. In Fig.~\ref{Fig2} they are shown 
as a function of the in-medium antiproton kinetic energy 
\begin{align}
E_{kin}=E_{\overline{p}}-m=
\sqrt{\vec{p\,}^{2}+\overline{m}^{*2}}+\overline{\Sigma}^{0}_{vp}-m
\end{align}
at three baryon densities (note that $\overline{\Sigma}^{0}_{vp}$ is 
negative for antinucleons by neglecting the isovector contributions, 
see Eq.~(\ref{Sigmas_nm})). 
The conventional Walecka approach (NL3 parametrization~\cite{NL3}, dashed 
curves) shows 
the usual linear behavior in density for the vector self-energy, while the scalar 
component saturates with density. Obviously, the Walecka self-energies are flat 
as a function of the kinetic energy. On the other hand, the NLD relativistic 
self-energies (solid curves) decrease 
with increasing kinetic energy at each baryon density. The strong suppression 
of the high-density and high-momentum tails of the NLD self-energies result from the 
momentum dependence of the fields, see Eqs.~(\ref{Sigmav_nm},\ref{Sigmas_nm}). 
In particular, at zero kinetic energy the Walecka model 
leads to a value of 
$\overline{\Sigma}_{vp}-\overline{\Sigma}_{s}\approx -700$~MeV, 
which remains constant in energy and is very deep. On the contrary, the NLD model 
reduces considerably the deepness of the in-medium antinucleon potential in 
all ranges of shown kinetic energies.

As found in Ref.~\cite{larionov}, in order to reproduce the data from antiproton-induced 
reactions within a transport model, the antinucleon-meson couplings of the Walecka 
model had to be rescaled 
by a phenomenological factor of $0.2-0.3$. In Fig.~\ref{Fig2} we show 
such calculations in 
the Walecka model with rescaled couplings 
(factor 0.25, dash-dotted curves in Fig.~\ref{Fig2}). As one can see, the rescaled Walecka 
model~\cite{larionov}, which is supposed to explain the antiproton-nucleus data, just 
reproduces in average the NLD results.

\begin{figure}[t]
\begin{center}
\unitlength1cm
\begin{picture}(12.,10.0)
\put(-1.0,0.0){
\makebox{\includegraphics[clip=true,width=0.85\columnwidth,angle=0.]
		 {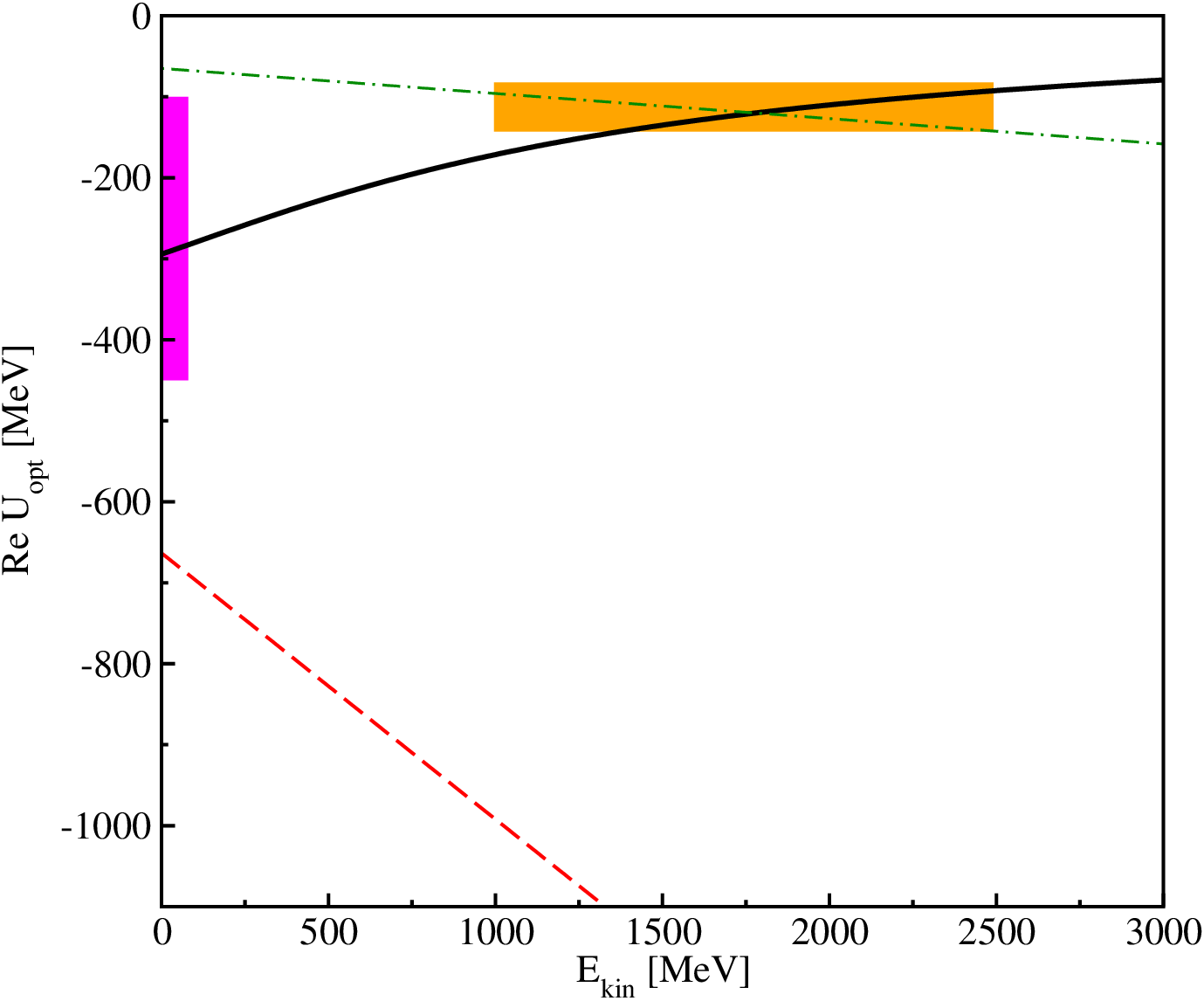}
		 }
}
\put(5.,1.5){
\makebox{\includegraphics[clip=true,width=0.45\columnwidth,angle=0.]
		 {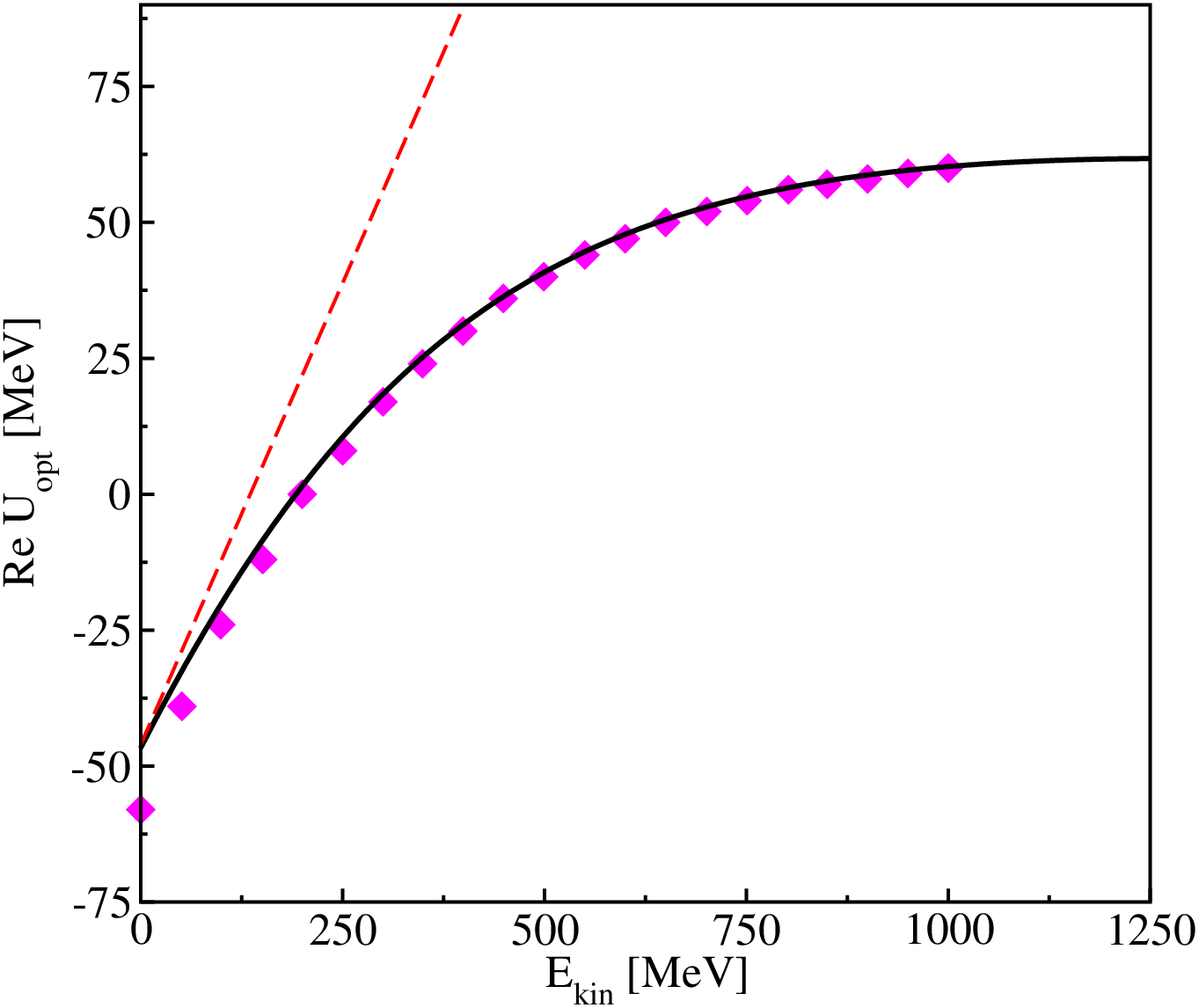}
		 }
}
\end{picture}
\caption{Real part of the Schr\"{o}dinger-equivalent optical potentials for 
antiprotons (main panel) and protons (insert) interacting with 
nuclear matter at saturation density as function of the kinetic energy. The 
shaded areas in the main panel and the filled circles in the inserted 
are extractions from empirical analyses~\cite{larionov,hama,data1}. 
The dashed (dot-dashed) curves are calculations in 
the (rescaled) Walecka model and the solid curves are calculations 
within the NLD approach.
}
\label{Fig3}
\end{center}
\end{figure}
The momentum dependence of the relativistic mean-fields can be probed by looking at the 
Schr\"{o}dinger-equivalent optical potential. Its real part is given by
\begin{equation}
 \mathfrak{Re} U_{\rm opt} = \frac{E}{m} V - S
 + \frac{1}{2m} \left( S^{2} - V^{2}\right)
\:, \label{U_opt}
\end{equation}
where $E=E_{N_{i}}$ ($E_{\overline{N}_{i}}$) is the energy of a 
nucleon (antinucleon) with bare mass $m$ inside nuclear matter at a fixed 
baryon density, $V=\Sigma_{vi}^{0}$ ($\overline{\Sigma}_{vi}^{0}$) is the 
corresponding vector self-energy and $S=\Sigma_{s}$ ($\overline{\Sigma}_{s}$) 
denotes the scalar self-energy component. 

Fig.~\ref{Fig3} shows $\mathfrak{Re} U_{\rm opt}$ as function of the kinetic 
energy for in-medium antiprotons at saturation 
density. The shaded areas for antiprotons around zero kinetic 
energy and antiprotons at high 
energies are the results of phenomenological analyses of Refs.~\cite{data1,larionov}. 
It can be seen that the momentum dependence is crucial for a correct 
description of the deepness of the optical potential (solid curve), whereas 
the standard Walecka approach (dashed curve) does not reproduce the absolute 
value and the trend of empirical data. The comparison with the empirical 
analyses is improved in the rescaled Walecka model (dot-dashed curve).

The inner graph in Fig.~\ref{Fig3} shows the corresponding results for 
the proton case, where the NLD calculations 
(solid curve) describe the Dirac phenomenology very well. However, the Walecka 
model (dashed curve) again disagrees with the data (filled symbols~\cite{hama}). 
We stress again that the NLD model parameters are kept fixed when going from 
in-medium proton to in-medium antiproton interactions. 

\begin{figure}[t]
\begin{center}
\unitlength1cm
\begin{picture}(12.,10.0)
\put(-1.,0.0){
\makebox{\includegraphics[clip=true,width=0.85\columnwidth,angle=0.]
		 {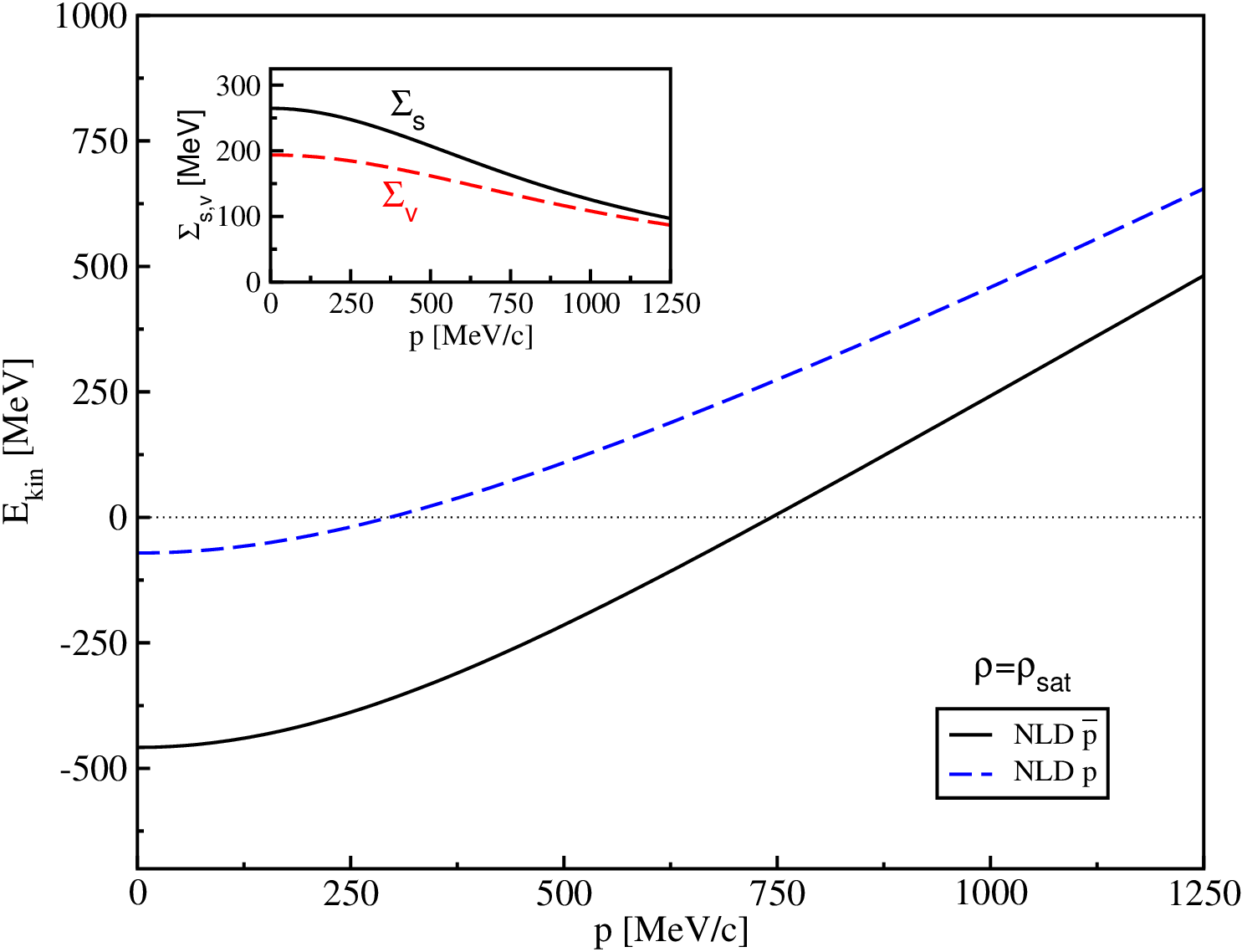}
		 }
}
\end{picture}
\caption{In-medium kinetic energy of a proton (dashed curve, NLD p) and an 
antiproton (solid curve, NLD $\bar{p}$) as function of single-particle momentum 
for nuclear matter at saturation density $\rho_{sat}=0.155~\fm^{-3}$. The 
calculations are performed in the NLD model. The inserted panel shows the 
momentum dependence of the scalar (solid curve) and vector (dashed curve) 
self-energies of a proton in nuclear matter at $\rho_{sat}$. 
}
\label{Fig4}
\end{center}
\end{figure}

At this point we would like to explain the behavior of the optical 
potential in the NLD model at kinetic energies around $E_{\rm kin}\simeq 0~\MeV$. 
This can be done by considering the in-medium dispersion 
relation, that is 
\begin{align}
m = \sqrt{p^{2}+\overline{m}^{*2}}+\overline{\Sigma}^{0}_{vp}
\end{align}
where $p=|\vec{p\,}|$ and is equal to
\begin{align}
p^2 = -(\overline{\Sigma}_{s})^2 + (\overline{\Sigma}^{0}_{vp})^{2}
+2m (\overline{\Sigma}_{s} - \overline{\Sigma}^{0}_{vp})
\label{ekinstopped}
\end{align}
with $\overline{\Sigma}_{s}$ ($\overline{\Sigma}^{0}_{vp}$) 
being the scalar (vector) RMF self-energies 
of the antiproton at fixed nuclear matter density. 
Hence, neglecting isovector contributions the single-antiparticle momentum in 
nuclear matter contains the sum of the scalar and the vector 
components of the RMF self-energies, 
$p \simeq \overline{\Sigma}_{s} - \overline{\Sigma}^{0}_{vp}$. 
Note, that in the case 
of in-medium protons the sign of the vector self-energy changes, 
$p \simeq \Sigma_{s} - \Sigma^{0}_{vp}$. 
In the antiproton case the single-antiparticle momentum is very big, while in the 
proton case it is essentially given by the small difference between scalar and 
vector mean-fields. This is demonstrated in Fig.~\ref{Fig4}, where 
the in-medium kinetic energy $E-m$ is shown as a function of 
the single-(anti)particle momentum for nuclear matter at saturation density 
of $\rho_{sat}= 0.155~\fm^{-3}$. Indeed, while in the proton case the momentum 
for a particle with in-medium kinetic energy $E_{kin}=0~\MeV$ is of the order 
of the Fermi momentum only, its value increases largely for the antiproton case. 
For an antiproton in matter at saturation density it approaches values of 
several hundreds of $\MeV$, namely $p(E_{kin}=0)\simeq 750~\MeV$, which are 
comparable with the nucleon mass. 


This momentum-shift shown in Fig.~\ref{Fig4} has significant 
consequences for the behavior of the potential deepness 
at zero in-medium kinetic 
energy, $\mathfrak{Re}U_{opt}(E_{kin}=0)$, when the self-energies depend 
on single-particle momentum. In fact, as it is shown in the 
inserted graph in Fig.~\ref{Fig4}, the NLD self-energies are reduced with 
increasing single-particle momenta. At a momentum 
of $p\simeq 750~\MeV$ the values of 
$\Sigma_{s}=163~\MeV$  and $\Sigma_{vp}=134~\MeV$ for the scalar 
$S$ and vector $V$ self-energies, respectively, are obtained. The sum 
of $\Sigma_{s}+\Sigma_{vp}=297~\MeV$ is considerably smaller than the 
corresponding sum at 
Fermi momentum. Since with increasing momentum the NLD fields are getting 
reduced, and in the antinucleon-case the sum of scalar and vector self-energies 
matters, therefore these effects results in the reduced value of the 
in-medium antinucleon potential.  


\begin{figure}[t]
\begin{center}
\unitlength1cm
\begin{picture}(12.,8.0)
\put(-1.3,0.0){
\makebox{\includegraphics[clip=true,width=0.9\columnwidth,angle=0.]
		 {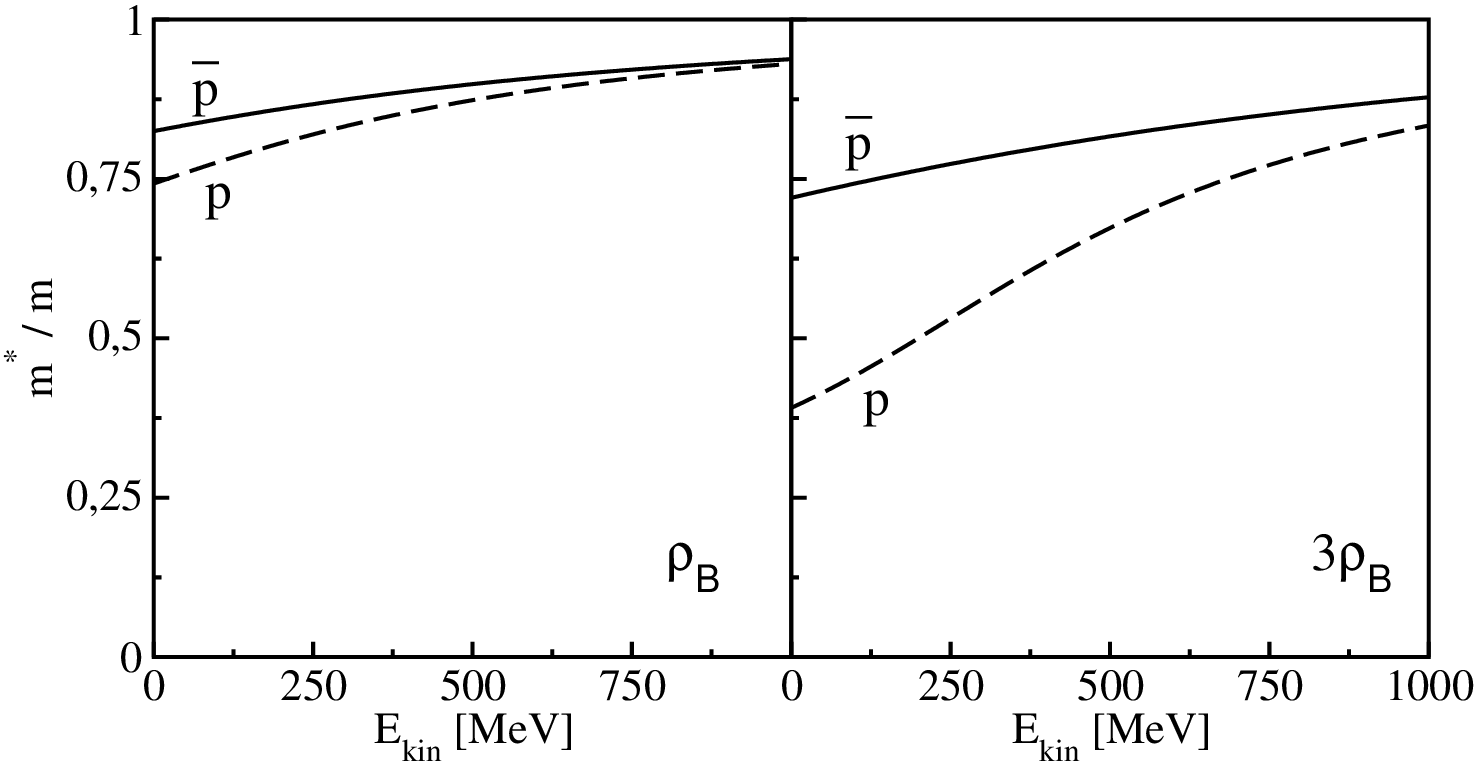}
		 }
}
\end{picture}
\caption{Effective mass of a proton (dashed curves) and an antiproton (solid curves) 
as function of the kinetic energy in nuclear matter at $\rho_{B}=\rho_{sat}$ (left) 
and at $\rho_{B}=3\rho_{sat}$ (right) in the NLD model.
}
\label{Fig5}
\end{center}
\end{figure}



\begin{figure}[t]
\begin{center}
\unitlength1cm
\begin{picture}(12.,10.0)
\put(-0.5,0.0){
\makebox{\includegraphics[clip=true,width=0.8\columnwidth,angle=0.]
		 {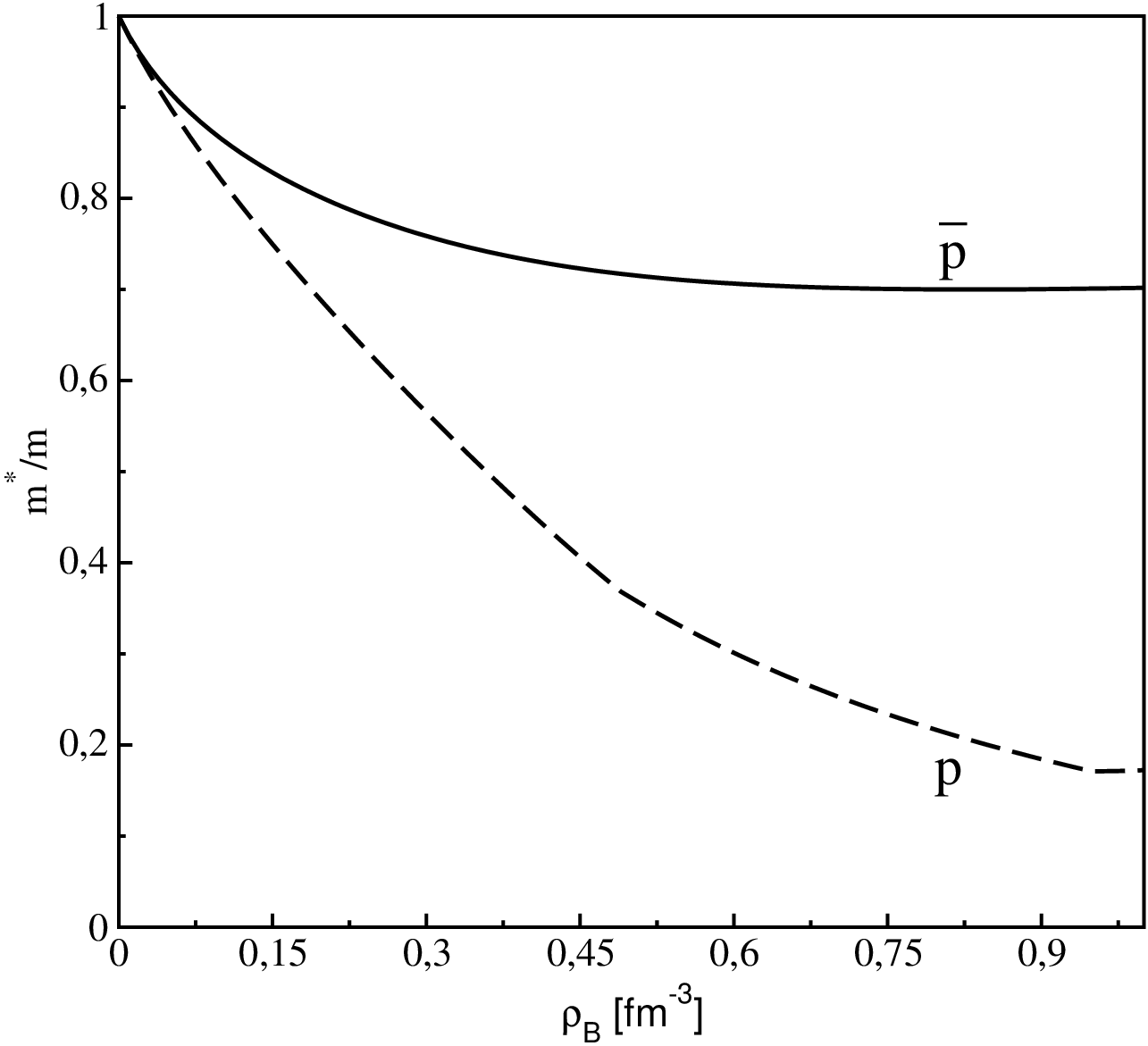}
		 }
}
\end{picture}
\caption{Effective mass for a proton (dashed curve) and for an antiproton (solid curve) 
at $E_{\rm kin}=0~\MeV$ in nuclear matter as a function of the baryon density 
in the NLD model.
}
\label{Fig6}
\end{center}
\end{figure}


Another effect generated by the momentum shift between in-medium 
protons and in-medium antiprotons, concerns the effective particle masses. 
Fig.~\ref{Fig5} shows the effective masses for protons, $m^{*}=m-\Sigma_{s}$, 
and antiprotons, $\overline{m}^{*}=m-\overline{\Sigma}_{s}$, in nuclear matter 
versus their kinetic energies and at two baryon densities. 
A mass shift between protons and antiprotons is observed, which becomes pronounced 
with decreasing kinetic energy and increasing baryon density. 
At the same value of the kinetic energy the proton scalar self-energy 
$\overline{\Sigma}_{s}$ 
is obtained at a smaller proton momentum than in the antiproton case, 
where $\overline{\Sigma}_{s}$ is 
larger. Therefore, at the same kinetic energy the effective proton mass 
is smaller as compared to the effective antiproton mass. This mass splitting between 
protons and antiprotons becomes moderate with increasing kinetic energy because of the 
less pronounced momentum dependence of $\overline{\Sigma}_{s}$ at high particle 
momenta. However, the 
mass splitting is strongly density dependent. This is demonstrated in Fig.~\ref{Fig6}, 
where the effective proton and antiproton masses are shown at zero kinetic energy 
as a function of the baryon density. 

One would expect that the only effect of the G-parity is the sign change 
in the vector self-energy. This is indeed the case in the standard Walecka 
models where there is no difference between the in-medium proton and antiproton 
masses. However, the NLD self-energies depend on momentum, as supported by 
Dirac phenomenologies. The impact of momentum dependence 
supplemented by different momentum shifts of protons and antiprotons in the 
scalar self-energies produces this observed mass splitting. 
Note the analogy with the isospin induced mass splitting between 
protons and neutrons in isospin-asymmetric nuclear matter. In the NLD model 
there is a mass splitting effect between nucleons and antinucleons 
also in G-asymmetric nuclear matter.

Next, we compare the NLD calculations with scattering data. This requires the 
knowledge of both, the real and the imaginary part of the in-medium 
optical potential. Using the dispersion relation one can calculate the imaginary 
part of $U_{opt}$ from the real part in Eq.~(\ref{U_opt}). 
The imaginary part of the Schr\"{o}dinger 
equivalent optical potential is given by~\cite{Disp}
\begin{equation}
\mathfrak{Im}U_{\rm opt}(p) = -\frac{2p}{\pi} \; {\cal P} \int_{0}^{\infty} 
 \frac{\mathfrak{Re}U_{\rm opt}(p^{\prime \;})}{p^{\prime\;2}-p^{2}} dp^{\prime}
\:, \label{DispRel}
\end{equation}
where $p\equiv |\vec{p}\;|$ stands for the antiparticle momentum and 
${\cal P}$ denotes the principal value. The real part 
$\mathfrak{Re}U_{\rm opt}$ is taken from the NLD model. Note that 
Eq.~(\ref{DispRel}) makes sense only if $\mathfrak{Re}U_{\rm opt}$ 
does not diverge with increasing momentum like in the standard Walecka model. 
This property of $\mathfrak{Re}U_{\rm opt}$ is realized in the NLD model.


\begin{figure}[t]
\begin{center}
\unitlength1cm
\begin{picture}(12.,10.0)
\put(-1.0,0.0){
\makebox{\includegraphics[clip=true,width=0.85\columnwidth,angle=0.]
		 {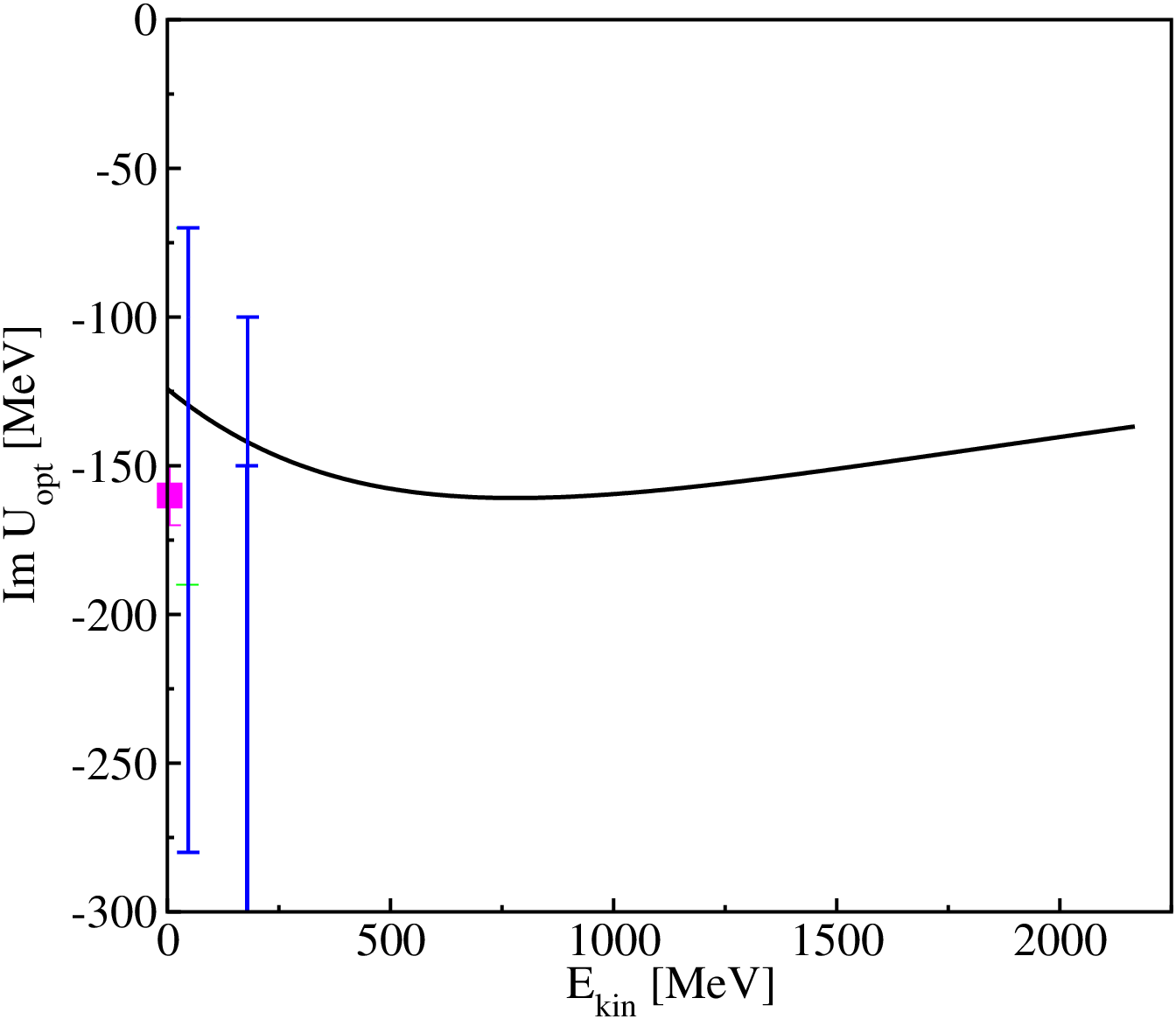}
		 }
}
\end{picture}
\caption{Imaginary part of the Schr\"odinger-equivalent optical 
potential (see Eq.~(\ref{DispRel})) as function of antiproton kinetic 
energy at saturation density. NLD calculations (solid curve) is compared with 
empirical extrapolations (symbol, vertical lines) taken from~\cite{data2}. 
}
\label{Fig7}
\end{center}
\end{figure}

The imaginary part of the in-medium antiproton optical potential is shown 
in Fig.~\ref{Fig7} as function of the antiproton kinetic energy. 
The empirical data are taken from the analyses on stopped antiprotons, 
see for details Refs.~\cite{data2}. The NLD calculations are in agreement 
with the empirical analyses within their uncertainties. The depth of 
the imaginary part of the optical potential is also consistent with 
phenomenological studies of antiprotonic atoms~\cite{data2}. 


\begin{figure}[t]
\begin{center}
\unitlength1cm
\begin{picture}(12.,10.0)
\put(-1.0,0.0){
\makebox{\includegraphics[clip=true,width=0.85\columnwidth,angle=0.]
		 {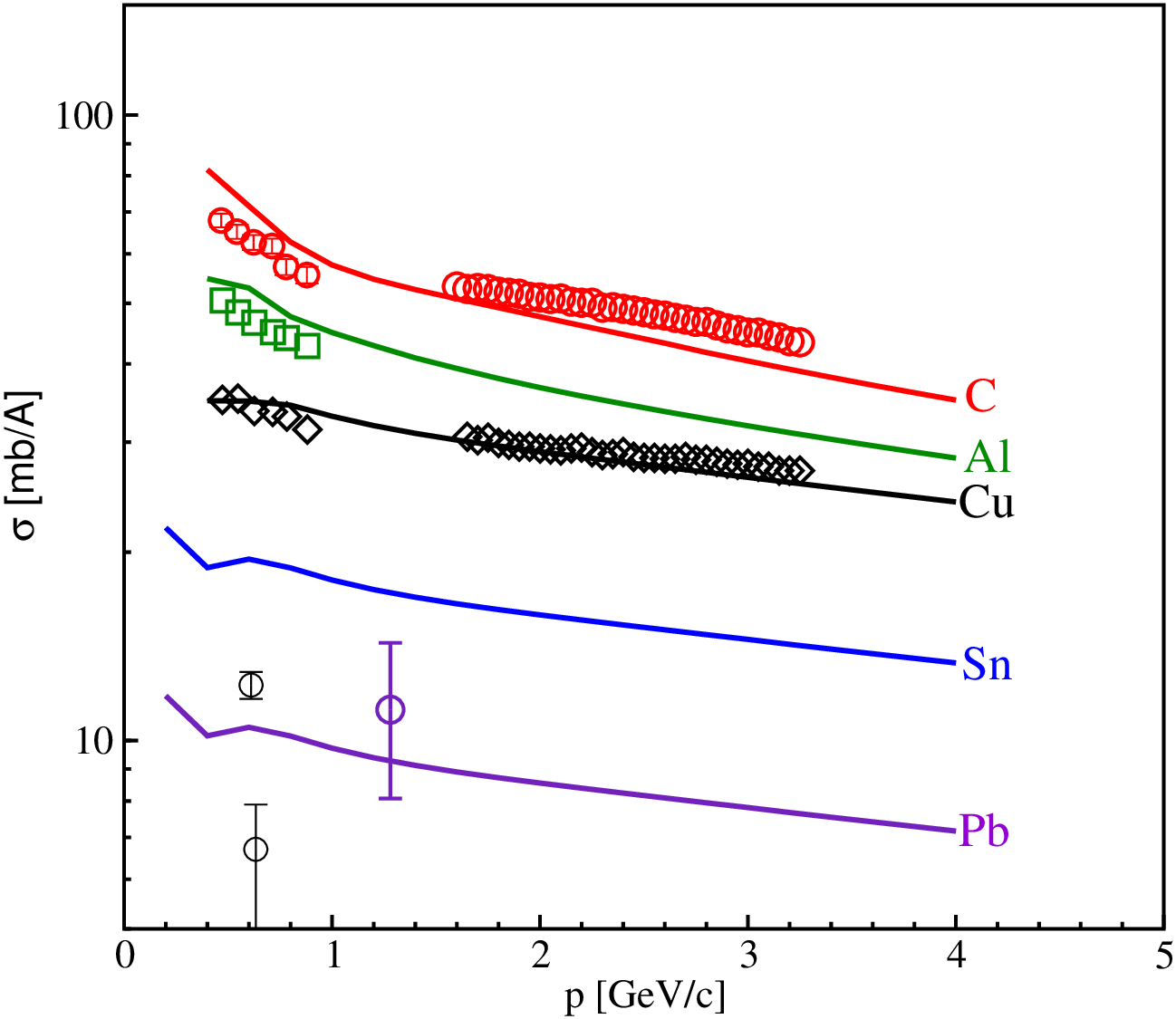}
		 }
}
\end{picture}
\caption{Total antiproton reaction cross sections as function of beam momentum 
for various target nuclei, as indicated. The NLD calculations (solid curves) are 
compared with experimental data (symbols), taken from~\cite{exp1,exp2,exp3,exp4} 
(see text for more details). 
}
\label{Fig8}
\end{center}
\end{figure}


Having both, the real and imaginary parts of the in-medium potential as function 
of beam momentum and density, we can now use the NLD optical potential to 
estimate the antiproton total cross sections in the scattering off nuclei. 
The numerical procedure is as follows. 
For a given nuclear target we use the empirical Wood-Saxon density 
distributions~\cite{WSdata}. They provide us with the density profiles 
which are used as an input to calculate the corresponding 
nuclear potential $U_{\rm opt}$ as a function of the radial coordinate. Then, 
at intermediate energies the scattering amplitude $f$ is obtained 
using the the eikonal approximation, that is 
\begin{align}
f(\vartheta) = ip \int_{0}^{\infty} db b J_{0}(\Delta b) (1 - e^{i\chi(b)}) 
\,.
\end{align}
The momentum transfer is given by $\Delta=2p\sin{\vartheta/2}$, 
$J_{0}$ denotes the Bessel function, 
$\vec{r}=\vec{b}_{\perp}+z\hat{n}$ and the eikonal $\chi$ reads 
\begin{align}
\chi = -\frac{p}{2E} \int_{-\infty}^{+\infty} U_{\rm opt}(b,z) dz
\,.
\end{align}
Using the optical theorem the total cross section takes the form 
\begin{align}
\sigma_{tot} = \frac{4\pi}{p}\mathfrak{Im}f(\vartheta=0)
\label{sigtot}
\,,
\end{align}
where the imaginary part of the amplitude is given by
\begin{align}
\mathfrak{Im}f(\vartheta) =  p\int_{0}^{\infty} db b J_{0}(\Delta b) 
\left( 1-\cos{(\mathfrak{Re}\chi)} e^{-\mathfrak{Im}\chi} \right)
\label{Imf}
\,.
\end{align}

In Fig.~\ref{Fig8} we show the NLD calculations for the total cross section 
as function of the antiproton beam momentum on different nuclear targets. 
The total antiproton cross sections are obtained from experimental absorption 
cross section data including optical model analyses for the missing contributions 
to the total one. In particular, 
the data symbols for the nuclear targets $C$, $Cu$ and $Al$ at low beam 
momenta are taken from Ref.~\cite{exp1}, whereas the experimental cross 
sections at the higher energies for $C$ and $Cu$ are from~\cite{exp2}. 
For antiproton reactions on the heavier $Pb$ target shown in Fig.~\ref{Fig8} 
there exist only few experimental estimations~\cite{exp3,exp4}. 
As one can see, 
the calculations reproduce the absolute values, momentum and mass number 
dependence of the total cross sections fairly well. 

The comparison of the NLD model with the results of empirical studies and 
with experimental data is remarkable. 
Our results compare well also with calculations done within 
the Giessen-BUU transport model~\cite{larionov}. In that work 
the imaginary part 
of the in-medium antiproton interaction was determined with the help of the 
collision integral of the transport equation, where the elastic, 
inelastic and annihilation processes were taken into account. 
There a standard RMF model has been also used. However, as we 
have already discussed, in Ref.~\cite{larionov} 
one had to strongly reduce the 
antinucleon-meson coupling constants to achieve a good description 
of antiproton absorption cross section data at low and intermediate energies. 
Therefore, in line with Ref.~\cite{larionov} a possible violation of G-parity was 
discussed. This conclusion is not supported by the results of the present work, 
where a simple constraint imposed by the energy dependence of the 
Schr\"{o}dinger-equivalent optical potential may result in a consistent G-parity 
conserved description of both, the in-medium nucleon and antinucleon interactions.

\section{Summary}

The NLD model with momentum dependent mean-fields has been applied 
in the past 
successfully for the descriptions of nuclear matter at low and high baryon densities. 
In this work the NLD 
formalism has been extended to the in-medium interactions of antinucleons. 
We studied the optical potential of antiprotons inside nuclear matter and 
then used it for the description of antiproton reactions off nuclei. 

It has been demonstrated that momentum-dependent interactions are able to 
describe simultaneously the nucleon and antinucleon optical potentials 
in agreement with available empirical data. In particular, we find 
a considerable reduction of the optical potential for antiprotons 
as compared to the standard RMF models. 
The same momentum dependence results in a new effective mass splitting 
between nucleons and antinucleons in nuclear matter. This sort of medium induced 
G-asymmetry becomes pronounced for low energy antinucleons and increases 
with rising baryon density.

Furthermore, the imaginary part of the in-medium optical potential has been 
determined by using the dispersion relation. The results concerning the imaginary 
part of the antiproton potential are compatible with available 
phenomenological analyses. 

We then applied the NLD optical potential to the description of antiproton-induced 
reactions on nuclear targets. Our calculations reproduce fairly well the 
dependence of the total reaction cross sections on beam momentum and 
nuclear mass number. 

We demonstrated that momentum dependent interactions, as realized in NLD 
model, lead to a consistent description of empirical information and experimental 
data. Together 
with the successful application of NLD approach to low and high density nuclear 
matter, we conclude that, a proper relativistic mean-field formalism should 
account for the momentum dependence of fields. This is important for a correct 
RMF description of nuclear matter and finite systems. 

As an outlook, it would be interesting to apply the NLD in reactions with 
heavy-ion and antiproton beams within transport theoretical descriptions. 
In particular, the mass splitting between nucleons and 
antinucleons, as predicted in this work, is expected to show up in various 
observables. 
For instance, it is known that in- and out-of-plane collective flows as well as 
subthreshold particle production are sensitive to possible mass splitting effects. 
Thus, the investigations of compressed matter in reactions 
relevant for \panda at FAIR may provide interesting signals as probes of 
the NLD predictions.

\section*{Acknowledgments}
This work was partially supported by DFG. 


\end{document}